# Trust Management Model For Cloud Computing Environment

Somesh Kumar Prajapati*, Suvamoy Changder**, Anirban Sarkar***

*Department of Computer Applications, National Institute of Technology, Durgapur, India*
*someshprajapati@gmail.com\*, suvamoy.nitdgp@gmail.com\*\*, sarkar.anirban@gmail.com\*\*\**

*Abstract*—Software as a service or (SaaS) is a new software development and deployment paradigm over the cloud and offers Information Technology services dynamically as "on-demand" basis over the internet. Trust is one of the fundamental security concepts on storing and delivering such services. In general, trust factors are integrated into such existent security frameworks in order to add a security level to entities collaborations through the trust relationship. However, deploying trust factor in the secured cloud environment are more complex engineering task due to the existence of heterogeneous types of service providers and consumers. In this paper, a formal trust management model has been introduced to manage the trust and its properties for SaaS in cloud computing environment. The model is capable to represent the direct trust, recommended trust, reputation etc. formally. For the analysis of the trust properties in the cloud environment, the proposed approach estimates the trust value and uncertainty of each peer by computing decay function, number of positive interactions, reputation factor and satisfaction level for the collected information.

## I. INTRODUCTION

Cloud computing [1] is a pervasive paradigm, where large pools of systems are connected in private or public networks, to provide dynamically scalable infrastructure for application, data and file storage. Services are provided "on demand" basis to cloud users over high-speed internet within the "X as a service (XaaS)" computing framework where X is defined as "Infrastructure", "Platform", and "Software". Among these, through the SaaS layer, cloud users get their applications as on-demand basis over the internet. One of the key benefits of SaaS is the ability to deliver the technology needs of a business as a service. This layer enables control and compliance over the environment to share a single instance of software among the several consumers [25].Security is a key concern for SaaS in cloud computing, as users store and access confidential data to and from the cloud. Under this circumstance, traditional security mechanism based on registration, authentication and authorization were no longer suitable for cloud computing environment [2] [3].A service provider might be authenticated and authorized, but this does not ensure that it exercises its authorization in a way that is expected [4]. As trust has been regarded as more essential security relationship than authorization in demotic, there is great significance of research towards trust relationship and trust based security mechanism within cloud environment which is mostly like human society. However, there exists very few trust management models for SaaS in cloud computing environment.

The concept of trust, from the perspective of information security, will correspond to a set of relations among entities that participate in a behavioral process [5]. These relations universally involve two entities, the service provider is called *trustor* and the subject requiring access to the services of trustor is called *trustee*. Trust establishment are based on the knowledge or experiences collected from the previous interactions of entities. In general, if the interactions conform to the intention of trustors, then trust evaluation will be correspondingly high in perspective of trustors. Otherwise, it will be accordingly low. In [6] [7], trust evaluation and reasoning methods have been proposed using probability models. Those methods do not consider fuzziness of trust itself, and their reasoning is based on pure probability models. However, in those approaches, authors have not discussed regarding the fundamental rules those are used to follow by the proposed trust models. Moreover, the design of trust models is still at the empirical stage.

In this paper, a formal trust management model has been proposed for SaaS from the basic concepts of trust. In the proposed approach, besides the time based experience, reputation concepts also has been considered for calculating the direct trust of the service providers in cloud environment. In the model, it has been perceived that before accessing the services from SaaS, consumers will ensure the trustworthiness of relevant service providers of the cloud environment as they provide services with different level of access according to the service level agreements, data security, performance etc. The proposed trust model is more suitable for SaaS in the cloud environment. Further, the model is capable to update the recommended trust values dynamically for each entity of the cloud.

The rest of the paper is organized as follows. Section II introduces previous research on trust management model and their approaches. Section III describes trust basics and trust properties. Section IV presents proposed trust management model. Section V shows a small case study on distributed file sharing in cloud environment. Finally conclusions and future work are presented in the last section.

## II. RELATED RESEARCH

In present literatures, trust based on human notation is applied widely to cope with new security concern in cloud. In [16], Sun *etal.* Proposed a framework to quantitatively measure trust, trust propagation and defend trust evaluation systems against malicious attacks. According to these works, three trust management models emerged as, policy based, social based and reputation based models. Reputation based systems such as Eigen Rep [17], are based on measuring reputation. They evaluate the trust in the peer and the trust in the reliability of the resource. Beth et al. [7] proposed a trust model for distributed networks and has distinguished recommendation trust from direct trust and gave their formal representations along with the rules to derive relationships and algorithm to compute direct trust values. Li et al.[20] introduced a domain-based cloud trust model to solve security issues in cross-cloud environment. Dawei et al. [15] proposed space-variant evaluation method for calculating recommended trust and time-variant comprehensive evaluation method for expressing direct trust. However, it hasused time based forgetting function for direct trust which is not enough because reputation factor also affect direct trust. Moreover, they assumed that all entities are honest and not able to resist malicious recommendations. Wang*et al.* [21] proposed an evaluation approach of subjective trust based on subjective trust cloud and has combined expected value with hyper entropy of subjective trust cloud to evaluate the randomness and fuzziness of subjective reputation. M. Rajarajan *et al.* [23] have presented a trust model to support service providers to verify trustworthiness of infrastructure providers in cloud environments. The trust values are calculated based on an opinion model in terms of belief, disbelief, uncertainty and base rate. However, the approach is completely based on service level agreements only. Deno and Sun [18] [24] proposed a Probabilistic Trust Management in Pervasive Computing that take trust value as a probability and which has used to device the satisfactory levels for interactions with its neighbor. But the approach lacks from the capability to distinguish between getting one positive outcome out of 2 interactions and getting 100 positive outcomes out of 200 interactions because in both cases the probability is equal to 0.5.Kang*et al.* [19] proposed a trust model based on expected value, entropy, hyper entropy and definition of trust cloud. He *et al.* [22] proposed trust model that has taken uncertain of trust into account and describe the trust degree and trust uncertainty in cloud. But it has assumed in the model that all entities are honest which deny the real time situation.

## III. TRUST BASICS

More precisely, trust lifecycle having three activities constituting the basic steps: trust establishment, trust update and trust revocation. Generally, trust is divided into two classes: direct trust and recommended trust [8] [9][13].Also it involves with two different kinds of *entities*, trustee (Consumers) and trustor (Service Providers). Direct trust is the trust based on own experience with entity and if two entities have no direct interactions, then trust relationship is established by another entity's recommendation, called recommended trust, as represented in Figure 1.The main concepts to inherent Trust Management(TM) arenamely the concepts of trustor, trustee and trust model.

### A. Trust Definition

Defining the trust concept is a challenging task since it may have different applications which may cause divergence in terminology. In [10] [11] trust is defined as: "Generally, an entity can be said to 'trust' a second entity when the first entity makes the assumption that the second entity will behave exactly as the first entity expects".

Moreover, Theo and al. [12] specify that trust must be related to a given service and define trust of a party *A* in a party *B* for a service *X* as "the measurable belief of *A* in *B* will behave dependently for a specified period within a specific context".

A concrete and mathematical definition of trust that has been followed in this article is given by Diego Gambetta [14]: Trust (or, symmetrically, distrust) is a particular level of the subjective probability with which an agent assesses that another agent or group of agents will perform a particular action, both before he can monitor such action (or independently or his capacity ever be able to monitor it) and in a context in which it affects his own action.

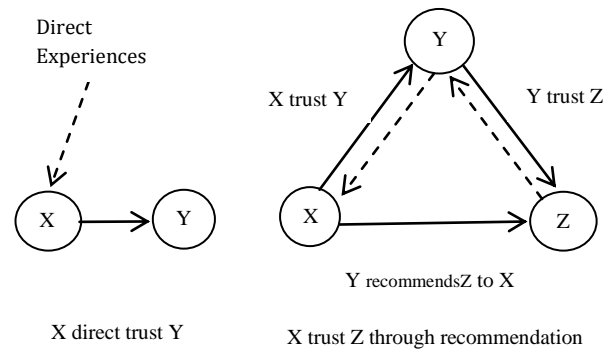

Figure 1. Direct trust and Recommended trust

### B. Trust Properties

Several trust properties were introduced in the literature [10] [15]. In this paper trust model is constructed based on several trust properties which is defined as follow:

*Asymmetry*: A trust relation is asymmetric. In fact, *X* trusting *Y* does not imply that *Y* trusts *X*, too.

*Reflexivity*: Trust is reflexive because each entity trusts itself.

*Context dependence*: A trust relation concerns a precise action on a precise object and cannot be generalized to other actions or objects.

*Scalability*: Trust is scalable since it may evolve during communication. This evolution implies trust level modification which also implies a modification of entities reputation. Trust levels precise trust degree while reputation designates the general appreciation of a given entity.

*Partial Transitivity*: Trust follows transitivity property.*Y* recommends *Z* to *X* if and only if *X* trusts *Y* and *Y* trusts *Z* otherwise, it's not follow transitivity property.

*Subjective*: Trust is a level of subjective probability.

*Uncertainty*: It is important characteristic of trust which means trust relationships between entities are fuzzy and stochastic, especially for stranger entities.

*Space based*: Recommendation trust satisfies the space-based variant property.

*Time based*: Direct trust satisfies the time-based variant property.

## IV. PROPOSED TRUST MANAGEMENT MODEL

Based on the trust basics in section III, a formal trust management model has been proposed for SaaS. Before accessing services and data from the cloud, consumers normally verify the trust level of service provider in cloud environment. So, this trust model first ensures the trustworthiness of service provider then access services and/or data. The main proposition in this paper is the modeling of a trust management system based on space variant evaluation for recommendation trust and time variant evaluation for direct trust. In addition, in the proposed model reputation concept also has been formalization.

### A. Proposed Trust Management Components

*(a) Trust:* Trust is a level of subjective probability hold by a trustor trusting a trustee, which is formed through the direct observation nature and/or recommendation from trusted entities, and depends on one or many performances of a trustee to fulfilling a particular service within a specific time and context. Trust is usually evaluated by trust degree and described with trust relation [15].

*(b) Trust Degree:* Trust degree $Td_{ij}$ is used to evaluate the degree of trust from a domain set of possible trust values that trustor $T_i$ in views trustee $T_j$ then the trust degree can be expressed as:

$$Td_{ij}(T_i, T_j, S_k, t) \quad \text{where } i \neq j;\ 0 \leq Td_{ij}(T_i, T_j, S_k, t) \leq 1$$

Where, $S_k$ is $k^{th}$ service and $t$ is defined as time. Trust degree has value between 0 and 1. Trust degree is calculated using direct trust $Td_{dir}$ or recommendation trust $Td_{recom}$ or if new entity joining a cloud environment first time then ignorance value $Td_{iv}$ is assigned.

$$\exists Td_{ij}(T_i,T_j,S_k,t) = \{ Td_{ij}(T_i,T_j,S_k,t) \rightarrow Td_{dir}(T_i,T_j,S_k,t)$$
$$\oplus Td_{recom}(T_i,T_j,S_k,t) \oplus Td_{iv}(T_i,T_j,S_k,t) \}$$

$$Td_{ij}(T_i,T_j,S_k,t) = \begin{cases} 0, & \text{if } n_d = 0, n_r = 0 \\ Dt(T_i,T_j,S_k,t), & \text{if } n_d \in \{1,2...\}, n_r = 0 \\ Rt(T_i,T_j,S_k,t), & \text{if } n_r \in \{1,2...\}, n_d = 0 \end{cases}$$

Where, $Dt(T_i, T_j, S_k, t)$ and $Rt(T_i, T_j, S_k, t)$ are the direct trust degree and recommendation trust degree of trustor $T_i$, in view of trustee $T_j$ about $k^{th}$ service $S_k$ at time $t$, $n_d$ and $n_r$ are the number of direct trust degree and recommendation trust degree.

*(c) Trust relation:* Trust relation $Tr_{ij}$ is the relationship between trustor $T_i$ and trustee $T_j$ from the trusted set $Q$ is described as a directed binary relation $Tr_{ij}<T_i,T_j> \in Q \times Q$. There are two type of trust relationship exists: one is direct trust and other is recommended trust.

*(d) Trust Levels:* Trust level represents the trustworthiness using degree of trust.

TABLE I. SATISFACTORY LEVEL

| Level | Label | Trustworthiness |
|---|---|---|
| I | No Opinion | $Td_{ij} = 0$ |
| II | Low distrust | $0 < Td_{ij} < 0.5$ |
| III | Medium trust | $Td_{ij} = 0.5$ |
| IV | High trust | $0.5 < Td_{ij} < 1$ |
| V | Complete trust | $Td_{ij} = 1$ |

*(e) Trust Chain:* The trust chain in cloud computing system is based on the partial transitive properties. If $\exists T_p, T_q, T_r, T_s, T_t \in Q, Tr_{pq}<T_p, T_q>, Tr_{qr}<T_q, T_r>$ then $T_p, T_q, T_r$ form a trust chain, denoted by $Tc_{pr}<<T_p, T_q, T_r>>$ and the length of the trust chain is 2. $Tr_{rs}<T_r, T_s>, Tr_{st}<T_s, T_t>$ then $T_r, T_s, T_t$ form a trust chain, denoted by $Tc_{rt}<<T_r,T_s, T_t>>$ and the length of the trust chain is 2. If both $Tc_{pr}<<T_p, T_q, T_r>>$ and $Tc_{rt}<<T_r, T_s, T_t>>$ chains are combined then the length of the trust chain is 4.

*(f) Trust Model:* In this paper trust management model is defined as: $TM = (T_i, T_j, Tr_{ij}, S_k, RF_i, t), i \neq j$; where trust model TM depends on trustor $T_i$, trustee $T_j$, trust relation between trustor and trustee is denoted as $Tr_{ij}$, $S_k$ is the $k^{th}$ service, $RF$ is trust reputation factor and time $t$.

*(g) Direct trust:* Direct trust is the trust relationship between two entities which have had direct interactions. In the model, each entity will maintain the trust values for all other entities in Direct Trust Table. The direct trust measure the subjective probability set about a trustee $T_j$ to the trustor $T_i$ in a specified service $S_k$ based on satisfaction level at time $t$ of the interaction. Trust degree is measured by direct experience of their interactions. Direct trust degree is denoted by $Dt(T_i,T_j, S_k,t), i \neq j$. Direct trust satisfies the time based variant property, means it depends on time based experience which is defined as:

*Time based experience:* Direct trust decay with time. The trust an entity has acquired at time $t$ in a perspective of a specified service might not be same as the trust attributed to him in the same perspective at time $t + \Delta t$

$$Dt(T_i, T_j, S_k, t + \Delta t) < Dt(T_i, T_j, S_k, t)$$

Let $t_c$ and $t_l$ denote the current time and last interaction time then decay function $\gamma$ is defined as

$$\gamma(t_c, t_l) = e^{-(\Delta t)^K} = e^{-(t_c - t_l)^K} \quad (1)$$

Where $K \in \{1, 2, 3...\}, \gamma(t_c, t_l) \in [0,1]$

$K$ determines the rate of decay of the direct trust degree with time $\Delta t$. If $RF_i$ is the reputation factor of trustor $i$ then calculate direct trust degree at present time $t_c$ using

$$Dt_{t_c}(T_i, T_j, S_k, t_{t_c}) = \frac{\sum_{l=1}^{N} \gamma(t_c, t_l) * Dt_{t_l}(T_i, T_j, S_k, t_{t_l})}{\sum_{l=1}^{N} \gamma(t_c, t_l)} + RF_i \quad (2)$$

*(h)Recommended trust:* If two entities have no direct interactions, then trust relationship is established by another entity's recommendation, called recommended trust. Trust degree is measured by another entity's evaluation results, as represented in Figure 1. In the model, each entity will maintain the list of all other entities with similar services and called as *Recommended List Table*. If for any trustor entity the direct trust value is not available then the recommended trust values will be updated dynamically in the Recommended List Table of trustee. Recommended trust measure the subjective probability set of a recommender about the trustee $T_j$ to the trustor $T_i$ in a specified service $S_k$ by one or many trust chains and is denoted by $Rt(T_i, T_j, S_k, t)$, $i \neq j$. Recommended trust of $r^{th}$ trust chain is calculated using,

$$Rt_r(T_i, T_j, S_k, t) = \frac{\sum_{m=1}^{length} W_{m,m+1} \cdot Dt_r(T_m, T_{m+1}, t)}{\sum_{m=1}^{length} W_{m,m+1}} \quad (3)$$

The weight between the node *m* and *m+1* in the $r^{th}$ trust chain $Tc_{ij}(r)$ is defined as:

$$W_{m,m+1} = \frac{n_p * sl}{n}, \quad W_{m,m+1} \epsilon [0,1] \quad (4)$$

where, $n_p$ is the number of positive interaction between node *m* and *m+1*, *n* is total no of interaction which is the summation of number of positive $n_p$ and negative $n_n$ interaction; *sl* is defined as satisfaction level. *sl* is depends on availability, processing capacity, recovery time, connectivity and peak-load performance which is defined in service level agreement, $sl \epsilon [0,1]$.

## V. CASE STUDY

In this section, a case study related to the distributed file sharing service has been represented under SaaS in cloud environment and trustworthiness of the related entities have been evaluated based on the proposed trust management model. In cloud environment, let a specific service of distributing files sharing, where the files have a desired distribution and availability. When any entity want to share file in cloud environment then first it need to ensure that whether a node or entity is trustworthy or not. The trustworthiness can be decided based on service level agreement(SLA) like processing capacity, recovery time, connectivity, peak-load performance and availability. For this case study, the trust management model $TM = (T_i, T_j, Tr_{ij}, S_k, RF_i, t)$ can be expressed using an example:

$TM = (T_i(Company\ i), T_j(Consumer\ j), Tr_{ij}$ (Direct, Recommended), $S_k$ (File sharing), $RF_i$ (High, Medium, Low), t (time))

In this TM model, let service provided by company *i*, service accessed by consumer *j*, $Tr_{ij}$ is the relationship between service provider and consumer either direct or recommended, file sharing service $S_k$, $RF_i$ is the reputation factor of service provider which is either high, medium or low and time is denoted by *t*. If a customer has access to a storage space in a cloud environment, it still has no selection criterion to determine to which cloud entity it will send a particular file. When an entity wants to share files with other entities, it will select trusted entities to store this file according to service level agreement.

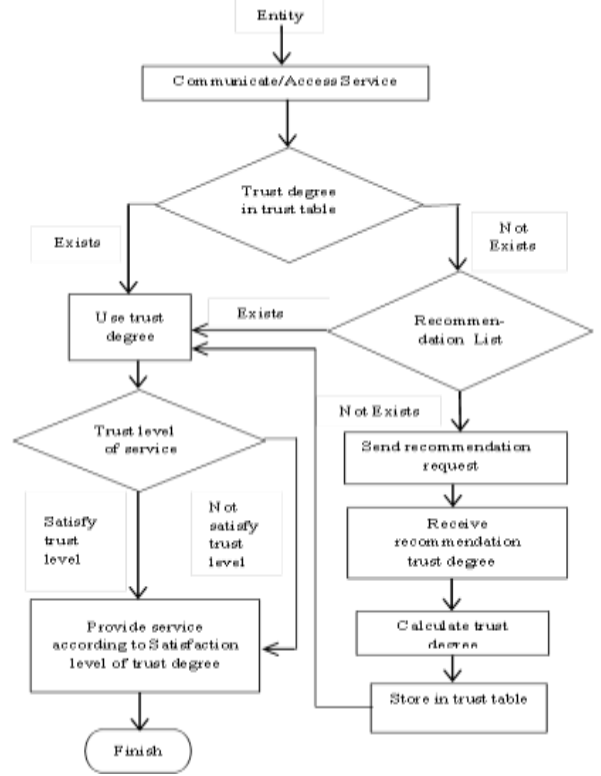

Figure 2. Trust management in file distribution

The trust relation is established using trust degree based on request sent to other entities in the cloud. Each entity will maintain two trust tables: direct trust table and the recommended list table. If an entity wants to calculate the trust degree of another entity then it first checks the direct trust table. If the trust degree value for the entity exists then it will check of last interaction time and then calculate the decay function using Eq.(1). After calculating decay function, direct trust for current time can be calculated using Eq.(2) and where reputation factor also will be considered. The direct trust for current time can be incremented or decremented according to decay function effect.

If this value is not available yet, then the recommended lists are checked to find an entity that has a direct trust relationship with the desired entity. In that case, the direct trust degree from this entity's direct trust table is used. If there is no value in table, then it sends a recommendation request to the other entities. The trust degree can be calculated based on the received recommendation trust degree response using Eq.(3).

Weight factor between entity *m* and *m+1* in trust chain can be calculated using Eq.(4) which is depends on number of positive interaction and satisfaction level of service level agreement and then store in table. The requesting entities will assign a greater trust degree to entities that having greater positive interaction out of total interactions with greater satisfaction level according to service level agreement. The calculated trust degree can be stored in trust table by either using direct trust or recommended trust and then can be checked the trust levels of intended services. In this proposed trust management model the trust level is defined in Table - I. As every service having different trust level so if any entity want to access service then entity must satisfy the trust level for particular service otherwise entity not allow to access the services.

## VI. Conclusion

This paper presents a formal trust management model based on the basics of the trust characteristics. The proposed model is capable to handle various cloud services access scenarios where entity has a past experience with the service or a stranger entity requesting to access the service without any identity or past interaction with the service. The work in this paper has defined the direct trust with a time-variant evaluation method and the recommended trust with a space-variant evaluation method. Motivated by human nature, the model also has considered the reputation factor of trustor to calculate the direct trust. The proposed approach also has used the satisfaction level to calculate recommended trust which is depends on service level agreements of the services resides in the cloud environment.

The future work will include the development of separate algorithms for evaluation of direct trust and recommendation trust schemes proposed in this model. Enhancing the proposed trust model towards more robust to resist malicious recommendations more rigorously also will be a prime focus in future work.